\documentclass[onecolumn]{article}

\usepackage{PRIMEarxiv}
\usepackage{algorithm}
\usepackage[utf8]{inputenc} 
\usepackage[T1]{fontenc}    
\usepackage{hyperref}       
\usepackage{url}            
\usepackage{booktabs}       
\usepackage{amsmath,amssymb,amsfonts}       
\usepackage{nicefrac}       
\usepackage{microtype}      
\usepackage{lipsum}
\usepackage{fancyhdr}       
\usepackage{graphicx}       
\graphicspath{{media/}}     
\usepackage{cite}
\usepackage{algorithmic}
\usepackage{graphicx}
\usepackage{textcomp}

\usepackage{amsthm}
\theoremstyle{definition}
\newtheorem{thm}{Theorem}
\newtheorem*{thm*}{Theorem}

\newtheorem*{lem*}{Lemma}
\newtheorem{pro}{Proof of Theorem}
\newtheorem*{pro*}{Proof}
\newtheorem{cor}{Corollary}
\newtheorem*{cor*}{Corollary}

\newtheorem*{Cor_pro*}{Proof}

\usepackage[subrefformat=parens]{subcaption}
\title{Pairing optimization via statistics: Algebraic structure in pairing problems and its application to performance enhancement\footnote{This research was funded in part by the Japan Science and Technology Agency through the Core Research for Evolutionary Science and Technology (CREST) Project (JPMJCR17N2), and in part by the Japan Society for the Promotion of Science through the
Grants-in-Aid for Scientific Research (A) (JP20H00233) and Transformative Research Areas (A) (JP22H05197). AR is a JSPS International Research Fellow.}}
\date{}
\author{
Naoki Fujita\footnotemark[1]\footnote{Department of Information Physics and Computing, Graduate School of Information Science and Technology, The University of Tokyo, Hongo, Bunkyo-ku, Tokyo 113-8656, Japan}\and
Andr\'e R\"{o}hm\footnotemark[1]\and
Takatomo Mihana\footnotemark[1] \and
Ryoichi Horisaki\footnotemark[1]\and
Aohan Li\footnotemark[2]\footnote{Graduate School of Informatics and Engineering, The University of Electro-Communications,
1-5-1 Chofugaoka, Chofu-shi, Tokyo 182-8585, Japan.}\and
Mikio Hasegawa\footnotemark[3]\footnote{Department of Electrical Engineering, Graduate School of Engineering, Tokyo University of Science, 6-3-1 Niijuku, Katsushika-ku, Tokyo 125-8585, Japan}\and
Makoto Naruse\footnotemark[1]}

\begin{document}
\maketitle

\begin{abstract}
Fully pairing all elements of a set while attempting to maximize the total benefit is a combinatorically difficult problem. Such pairing problems naturally appear in various situations in science, technology, economics, and other fields. 
In our previous study, we proposed an efficient method to infer the underlying compatibilities among the entities, under the constraint that only the total compatibility is observable. 
Furthermore, by transforming the pairing problem into a traveling salesman problem with a multi-layer architecture, a pairing optimization algorithm was successfully demonstrated to derive a high-total-compatibility pairing. 
However, there is substantial room for further performance enhancement by further exploiting the underlying mathematical properties.
In this study, we prove the existence of algebraic structures in the pairing problem. We transform the initially estimated compatibility information into an equivalent form where the variance of the individual compatibilities is minimized. 
We then demonstrate that the total compatibility obtained when using the heuristic pairing algorithm on the transformed problem is significantly higher compared to the previous method. 
With this improved perspective on the pairing problem using fundamental mathematical properties, we can contribute to practical applications such as wireless communications beyond 5G, where efficient pairing is of critical importance.

\newline
\end{abstract}

\section{Introduction}
\label{Introduction}
The procedure of generating pairs of elements among all entries of a given system often arises in various situations in science, technology, and economy \cite{gale1962college, roth1982economics, ergin2017dual, kohl2004airline, gambetta2017building, gao20113d, bellur2007improved}. 
Here we call such a process pairing, and the number of elements is considered to be an even number for simplicity. 
One immediately obvious problem is that the number of pairing configurations grows rapidly with the number of elements. 
The number of possible pairings is given by $(n-1)!!$, where $n$ indicates the number of elements in the system and $!!$ is the double factorial operator. 
For example, when $n$ is 100, the total number of possible pairings is on the order of  $10^{78}$. 
Hence, finding the pairing that maximizes the benefit of the total system is difficult. 
Notably, the pairing problem corresponds to the Maximum Weighted Matching (MWM) problem on the complete graph \cite{edmonds1965paths, gabow1990data, huang2012efficient,pettie2012simple, cygan2015algorithmic, duan2010approximating, hanke2010new, duan2014linear}.

An example of a pairing problem is found in a recent communication technology called Non-Orthogonal Multiple Access (NOMA)  \cite{aldababsa2018tutorial, ding2015impact, chen2019proportional, ali2021optimizing, zhang2020energy, shahab2016user, zhu2018optimal}. 
In NOMA, multiple terminals simultaneously share a common frequency band to improve the efficiency of frequency usage. 
The simultaneous use of the same frequency band causes interference in the signals from the base station to each terminal. 
To overcome this problem, NOMA uses a signal processing method called Successive Interference Cancellation (SIC) \cite{higuchi2015non} to distinguish individual channel information in the power domain, allowing multiple terminals to rely on the same frequency band. For simplicity, here we consider that the number of terminals that can share a frequency is given by two. 
Herein, the usefulness of the whole system can be measured by the total communication quality, such as high data throughput and low error rate, which depends crucially on the method of pairing.

The most fundamental parameter of the pairing problem is the merit between any two given elements, which we call  individual compatibility, while the summation of compatibilities for a given pairing is called its total compatibility. 
The detailed definition is introduced below. 
Our goal is to derive pairings yielding high total compatibility. 

In general, we do not need to assume that the individual compatibility of a pair is observable, i.e., only the total compatibility of a given pairing may be observed. 
Our previous study \cite{fujita2022efficient} divided the pairing problem into two phases. 
The first is the observation phase, where we observe total compatibilities for several pairings and estimate the individual compatibilities. 
The second is the combining phase, in which a search is performed for a pairing that provides high total compatibility. This procedure is referred to as pairing optimization. The search is based on the compatibility information obtained in the first phase. 
In \cite{fujita2022efficient}, we show that the pairing optimization problem can be transformed into a travelling salesman problem (TSP) \cite{halim2019combinatorial} with a three-layer structure, allowing us to benefit from a variety of known heuristics. 

However, we consider that there is substantial room for further performance optimization. 
This study sheds new light on the pairing problem from two perspectives. 
The first is to clarify the algebraic structure of the pairing optimization problem. 
Because we care only about the total compatibility when all elements are paired, there are many compatibility matrices (defined in Section~\ref{Problem Setting}) that share the same total compatibilities. 
In other words, we can consider an equivalence class of compatibility matrices that yield the same total compatibilities and that cannot be distinguished if individual compatibilities are not measurable.
We show that the compatibility matrices in each equivalence class have an invariant value. 

Second, although any compatibility matrices in the same equivalence class theoretically provide the same total compatibility, the heuristic pairing optimization process can result in different total compatibility values. 
These differences are not caused by incomplete or noisy observations, but are due to the convergence properties of the heuristic pairing algorithms, which yield better results on some distributions than others.
We examine how the statistics of the compatibility matrix affect the pairing optimization problem and propose a compatibility matrix that yields higher total compatibility after optimization. 
More specifically, we propose a transformation to the compatibility matrix that minimizes the variance of the elements therein, which we call the variance optimization. 
We confirmed numerically that enhanced total compatibility is achieved via the compatibility matrix after variance optimization. Furthermore, the proposed variance optimization algorithm may also be applicable when no observation phase is required, i.e., when the individual compatibilities are directly observable. In other words, there are cases where a compatibility matrix unsuitable for a heuristic combining algorithm can be converted to one that is easily combinable.

The remainder of this paper is organized as follows. 
In Section \ref{Problem Setting}, we define the pairing optimization problem mathematically. 
Section \ref{Mathematical Property} describes the mathematical properties of the equivalence class. 
Section \ref{Variance Optimization} explains the concept of variance optimization and presents a solution by which it can be achieved. 
Section \ref{simulation} presents results of numerical simulations of the proposed variance optimization. 
Finally, Section \ref{Conclusion} concludes the paper.

\section{Problem Setting}
\label{Problem Setting}
In this section, we provide a mathematical definition of the pairing optimization problem that we address in this study, and define some of the mathematical symbols used in the following discussion. 
In addition, we explain the constraints applied to the pairing optimization problem.

\subsection{Pairing Optimization Problem}
Here we assume that the number of elements is an even natural integer $n$, while the index of each element is a natural number between 1 and $n$. 
Parts of the pairing problem can be described elegantly in set theory, while others benefit from using matrix representations. We will use either, where appropriate.
Here we use $\mathbb{U}(n)$ to denote the set of $n$ elements:
\begin{eqnarray}
    \mathbb{U}(n)\equiv \{i\mid i\in\mathbb{Z}, 1\leq i \leq n\}.
\end{eqnarray}
Then, we define the set of all possible pairs for $\mathbb{U}(n)$ as $\mathbb{P}(n)$, which contains $N(N-1)/2$ pairs:
\begin{eqnarray}
    \mathbb{P}(n)\equiv\{\{i,j\}\mid i,j\in\mathbb{U}(n), i<j\}.\\
\end{eqnarray}
To describe the compatibilities of these pairs, we now define a ``compatibility matrix’’ $C$ as follows:
\begin{eqnarray}
    &&C\in \mathbb{R}^{n \times n},\nonumber\\
    &&\forall \{i,j\}\in \mathbb{P}(n), C_{i,j}=C_{j,i}, \nonumber\\
    &&1\leq i \leq n, C_{i,i}=0.\nonumber
\end{eqnarray}
The compatibility between elements $i$ and $j$ is denoted by $C_{i,j}\in\mathbb{R}$. 
The matrix $C$ is always symmetric and the major diagonal is zero, because pairing $i$ and $j$ does not depend on the order of elements and an element cannot be paired with itself.
The set of all possible compatibility matrices is denoted as $\Omega_n$ when the number of elements is $n$. 
In other words, $\Omega_n$ is the set of all $n\times n$ symmetric distance matrices, or symmetric hollow matrices.
To describe a pairing, i.e., which elements are paired together, we now define a pairing matrix $S\in \mathbb{R}^{n \times n}$:
\begin{eqnarray*}
    &&\forall \{i,j\}\in \mathbb{P}(n), S_{i,j}=S_{j,i} \textrm{\, and \,}S_{i,j} \in\{0,1\} ,\\
    &&1\leq i \leq n, S_{i,i}=0 ,\\
    && \forall i, \sum_{j=1}^n S_{i,j}=1 .
\end{eqnarray*}
$S$ is symmetric, because pairing element $i$ with $j$ is equivalent to pairing $j$ with $i$. 
The pairing matrix $S$ is also hollow, because pairing $i$ with itself is not allowed.
Each row and column contains only a single non-zero element, as each element $i$ can only be paired once.
Therefore, a pairing matrix $S$ is an $n\times n$ symmetric and hollow permutation matrix.
We define the set of all pairing matrices $\mathbb{S}(n)\equiv\{S\}$ when the number of elements is $n$. 
\begin{eqnarray}
    S\in \mathbb{S}(n),
\end{eqnarray}
To derive the set representation of a pairing, we introduce the map $f_{\textrm{set}}$ as follows:
\begin{eqnarray}
    f_{\textrm{set}}(S)\equiv\{\{i,j\}\mid i<j\textrm{\, and \,}S_{i,j}=1\}.    
\end{eqnarray}    

A function denoted by $\langle X,C\rangle$ is then defined as follows, using the Frobenius inner product $\langle \cdot \rangle_{\textrm{F}}$:
\begin{eqnarray*}
    &&C\in\Omega_n,\nonumber\\
    &&X\in \mathbb{R}^{n \times n},\\
    &&\langle X,C\rangle=\frac{1}{2}\langle X,  C\rangle_{\textrm{F}}.
\end{eqnarray*}
For a given compatibility matrix $C$, we call $\langle S,C\rangle$ for $S\in\mathbb{S}(n)$ the ``total compatibility’’ for pairing $S$. 
This formulation is equivalent to the one used in our previous work \cite{fujita2022efficient}, and corresponds to summing the individual compatibilities $C_{i,j}$ of the pairs defined by~$S$:
\begin{eqnarray*}
    &&\langle S,C\rangle = \sum_{\{i,j\} \in f_{\textrm{set}}(S)} C_{i,j} 
\end{eqnarray*}

For any given compatibility matrix $C$, the pairing optimization problem can then be formulated as follows:
\begin{eqnarray*}
    &&\textrm{max:}\,\langle S,C\rangle,\nonumber\\
    &&\textrm{subject\,to:}\,S \in \mathbb{S}(n).\nonumber
\end{eqnarray*}

\subsection{Limited Observation Constraint}
As briefly mentioned in Section \ref{Introduction}, in practice there may often exist one more constraint on the pairing optimization problem. 
We will assume that initially we do not know each compatibility value. 
Moreover, we assume that only the value of total compatibility $\langle S,C\rangle$ for any pairing $S\in\mathbb{S}$ is observable. 
We call this condition the ``Limited Observation Constraint’’. 

Under this constraint, we must execute two phases, the ``Observation Phase’’ and the ``Combining Phase’’, as introduced in our previous study~\cite{fujita2022efficient}. 
First, we estimate the ground-truth compatibility matrix $C^g$ through observations of the total compatibilities of several pairings in the observation phase. 
We denote the estimated compatibility matrix by $C^e$. 
Our previous work \cite{fujita2022efficient} calculated the minimum number of observations that are necessary for deducing $C^e$ and presents a simple algorithm for doing so efficiently.

\section{Mathematical Properties of the Pairing Problem}
\label{Mathematical Property}
In this section, we consider algebraic structures in the pairing problem. 
An equivalence relation is defined among compatibility matrices to construct equivalence classes. 
Then we show a conserved quantity within the equivalence class and that all members of the class yield the same total compatibility for any given pairing. 
Furthermore, the statistical properties of compatibility matrices are examined, forming the mathematical foundation of the variance optimization to be discussed in Section \ref{Variance Optimization}.

\subsection{Adjacent Set}

\begin{figure}[h]
 \centering
 \includegraphics[width=9cm]{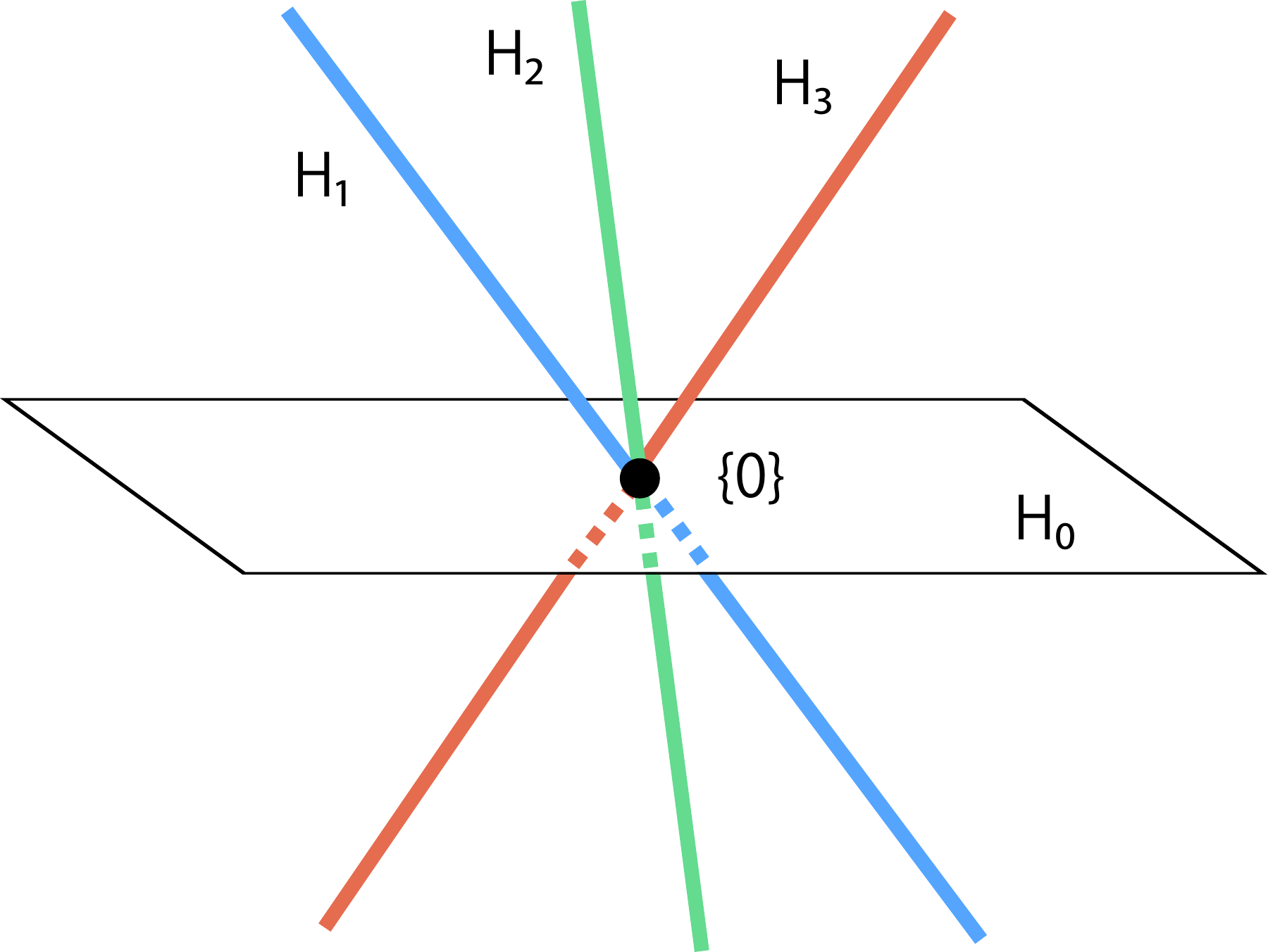}
 \caption{A schematic illustration of the relationship among $H_0, H_1, H_2$ and $H_3$.}
 \label{structure}
\end{figure}

We define the adjacent set matrix $R_i (1\leq i \leq n)$ as follows:
\begin{eqnarray}
    &&R_i\in \mathbb{R}^{n \times n},\nonumber\\
    &&(R_i)_{k,l}=\begin{cases}
        1\textrm{\quad if\quad}i\in\{k,l\}\textrm{\,and\,}k\neq l\\
        0\textrm{\quad otherwise\quad}.
    \end{cases}
\end{eqnarray}
We can also describe $f_{\textrm{set}}(R_i)$ as follows:
\begin{eqnarray}
    f_{\textrm{set}}(R_i)=\left\{\{i,j\}\mid 1\leq j \leq n, j\neq i\right\}.
\end{eqnarray}
With these adjacent sets, the following theorem holds. 
\begin{thm}
\label{thm_basis}
$C\in\Omega_n$ is fully determined by $\{\langle S,C\rangle\mid S\in\mathbb{S}(n)\}$ and $\{\langle R_i, C\rangle\mid 1\leq i\leq n-1\}$. 
\end{thm}
Note that $\langle R_n, C\rangle$ is not included, i.e., only $n-1$ terms involving $R_i$ are needed. Here, we have chosen to exclude index $n$ without loss of generality.
\begin{pro}
Our strategy to prove this involves calculating the dimension of the involved subspaces.
First, we prove the equation
\begin{eqnarray}
    \textrm{span} \{S\}_{S\in\mathbb{S}(n)}\cap\textrm{span} \{R_i\}_{1\leq i\leq n-1}=\{O_n\}
\end{eqnarray}
where $O_n$ denotes the $n\times n$ zero matrix.
Then, we focus on the following equation to check linear independence. Here, we number all pairings such as $S_1, S_2, \cdots S_u \cdots S_{(N-1)!!}$. We introduce the coefficients $a_u$ and $b_v$ and calculate the overlap of the spans:
\begin{eqnarray}
&&1\leq u \leq (n-1)!!, a_u\in\mathbb{R},\nonumber\\
&&1\leq v \leq n-1, b_v\in\mathbb{R},\nonumber\\
&&\sum_{u=1}^{(n-1)!!}a_u S_u=\sum_{v=1}^{n-1}b_v R_v.
\end{eqnarray}
We focus on the summation of the $k$th-column on both sides. Note that for every $S_u$ there is exactly one non-zero element in column $k$, while for $R_v$ there may be more than one if $v = k$ and $1\leq k\leq n-1$, or exactly one non-zero element otherwise. Then, the following equations hold:\\
When $1\leq k\leq n-1$
\begin{eqnarray}
    \label{thm1_3}(n-2)b_k+\sum_{l=1}^{n-1}b_l-\sum_{l=1}^{(n-1)!!}a_l=0.
\end{eqnarray}
When $k=n$ (because of our choice in formulating Theorem~\ref{thm_basis})
\begin{eqnarray}
    \label{thm1_4}\sum_{l=1}^{n-1}b_l-\sum_{l=1}^{(n-1)!!}a_l=0.
\end{eqnarray}
With Equations \eqref{thm1_3} and \eqref{thm1_4}, $b_k=0\,(1\leq k \leq n-1)$ holds. This means that
\begin{eqnarray}
    &&\textrm{span} \{S\}_{S\in\mathbb{S}(n)}\cap\textrm{span} \{R_i\}_{1\leq i\leq n-1}=\{O_n\},\label{thm1_hokuukan}\\
    &&\label{thm1_rest}\textrm{dim\,span} \{R_i\}_{1\leq i\leq n-1}=n-1.
\end{eqnarray}
By our previous study \cite{fujita2022efficient},
\begin{eqnarray}
    \label{thm1_previous}\textrm{dim\,span} \{S\}_{S\in\mathbb{S}(n)}=L_{\textrm{min}}(n).
\end{eqnarray}
Here, we denote $L_{\textrm{min}}(n)\equiv (n-1)(n-2)/2$.
By Equations \eqref{thm1_rest} and \eqref{thm1_previous}, the following equation holds:
\begin{eqnarray}
    \textrm{dim\,span} \{S\}_{S\in\mathbb{S}(n)}+\textrm{dim\,span} \{R_i\}_{1\leq i\leq n-1}=\textrm{dim\,} \Omega_n\label{thm1_dim_equal}.
\end{eqnarray}  
Therefore, by Equations \eqref{thm1_hokuukan} and \eqref{thm1_dim_equal},
\begin{eqnarray}
    \textrm{dim\,span} \{S\}_{S\in\mathbb{S}(n)}\cup\textrm{span} \{R_i\}_{1\leq i\leq n-1}=\textrm{dim\,} \Omega_n.\label{dim_same}
\end{eqnarray}
The pairing matrices $S$ are a subset of $\Omega_n$. In addition, the adjacent set matrices $R_i$ are also a subset of $\Omega_n$. Therefore, the following equation holds:
\begin{eqnarray}
    \textrm{span}\{S\}_{S\in\mathbb{S}(n)}\cup\textrm{span} \{R_i\}_{1\leq i\leq n-1}\subseteq \Omega_n.\label{space_same}
\end{eqnarray}
With Equations \eqref{dim_same} and \eqref{space_same},
\begin{eqnarray}
    \textrm{span}\{S\}_{S\in\mathbb{S}(n)}\cup\textrm{span} \{R_i\}_{1\leq i\leq n-1}= \Omega_n.
\end{eqnarray}
That is, $\{S\}_{S\in\mathbb{S}(n)}$ plus $\{R_i\}_{1\leq i\leq n-1}$ can construct $\Omega_n$. Finally, $\langle S,C\rangle$ is a linear transformation of $S$ which comes from the property of the Frobenius inner product. Therefore, $C\in\Omega_n$ can be constructed as a linear combination of $\{\langle S,C\rangle\mid S\in\mathbb{S}(n)\}$ and $\{\langle R_i, C\rangle\mid 1\leq i\leq n-1\}$. Therefore, the theorem holds. \qed
\end{pro}
\begin{cor}
\begin{eqnarray}
&& A, B\in\Omega_n,\nonumber\\
&& A= B\textrm{\quad if and only if \quad}\nonumber\\
&&\forall S\in\mathbb{S}(n), \nonumber\\
&&\langle S,A\rangle=\langle S,B\rangle\textrm{\,and\,}
1\leq i\leq n, \langle R_i, A\rangle=\langle R_i, B\rangle. \label{cor_1}
\end{eqnarray}
\end{cor}
This corollary is a special case of Theorem \ref{thm_basis} because Equation \eqref{cor_1} means that $A$ and $B$ have the same total compatibilities for all pairings and all adjacent sets.\\
Here, we present an example for Theorem \ref{thm_basis} for the $n=4$ case to illustrate the relationship of the involved subspaces. We define the following $H_i$:
\begin{eqnarray}
    H_i=\begin{cases}
        \textrm{span}\{S\}_{S\in\mathbb{S}(n)} \textrm{\quad if\quad} i=0,\\
        \textrm{span}\{R_i\} \textrm{\quad if\quad} 1\leq i\leq n-1.
    \end{cases}
\end{eqnarray}
We represent $H_i$ as follows where $D_{i,j}\in\Omega_n$ is defined as the $n \times n$ matrix whose $(i,j)$th element is 1 and all other elements are $0$:
\begin{eqnarray}
    &&H_i=\begin{cases}
        \textrm{\quad if\quad} i=0,\\
        \{k_1(D_{1,2}+D_{3,4})+k_2(D_{1,3}+D_{2,4})+k_3(D_{1,4}+D_{2,3})\mid k_1,k_2,k_3\in\mathbb{R}\} \\
        \textrm{\quad if\quad} i=1,\\
        \{k_4(D_{1,2}+D_{1,3}+D_{1,4})\mid k_4\in\mathbb{R}\}\\
        \textrm{\quad if\quad} i=2,\\
        \{k_5(D_{2,1}+D_{2,3}+D_{2,4})\mid k_5\in\mathbb{R}\}\\
        \textrm{\quad if\quad} i=3,\\
        \{k_6(D_{3,1}+D_{3,2}+D_{3,4})\mid k_6\in\mathbb{R}\},
    \end{cases}\\
    &&\bar{H}=\{l_{i,j}D_{i,j}\mid 1\leq i< j\leq n, l_{i,j}\in\mathbb{R}\}.
\end{eqnarray}
The image of these spaces is represented in Figure \ref{structure}. That is,
\begin{eqnarray}
    &&0\leq i< j\leq n-1, i\neq j, H_i\cap H_j=\{O_n\},\\
    &&\bar{H}=H_0\cup H_1\cup H_2\cup H_3.
\end{eqnarray}

\subsection{Equivalence Class}
We define the relation $\sim$ as follows:
\begin{eqnarray}
&& A, B\in\Omega_n,\nonumber\\
&& A\sim  B\textrm{\quad if and only if\quad}\forall S\in \mathbb{S}(n), \langle S,A\rangle=\langle S,B\rangle.
\end{eqnarray}
This represents an equivalence relationship between $ A$ and $ B$, leading to the construction of an equivalence class. 

Regarding this equivalence class, the following theorem holds:
\begin{thm}
\label{thm_equivalence}
\begin{eqnarray}
&& A, B\in\Omega_n, \nonumber\\
&& A\sim B\textrm{\quad if and only if\quad}\nonumber\\
&&\forall \{i,j\}\in\mathbb{P}(n),\nonumber\\
&&A_{i,j}-\frac{1}{n-2}\left(\langle R_i, A\rangle+\langle R_j, A\rangle\right)=B_{i,j}-\frac{1}{n-2}\left(\langle R_i, B\rangle+\langle R_j, B\rangle\right).
\end{eqnarray}
That is, for any matrix $C$ in the equivalence class, the values given by the following are conserved.
\begin{eqnarray}
    \forall \{i,j\}\in\mathbb{P}(n), C_{i,j}-\frac{1}{n-2}\left(\langle R_i, C\rangle+\langle R_j, C\rangle\right).
\end{eqnarray} 
\end{thm}
The matrix form of the conserved values is described in Appendix~\ref{Matrix Form of Conserved Quantities}.
\begin{pro}
First, we prove sufficiency. 
We assume that the following equation holds:
\begin{eqnarray}
\label{thm1_2}
\forall \{i,j\}\in \mathbb{P}(n), A_{i,j}-\frac{1}{n-2}\left(\langle R_i, A\rangle+\langle R_j, A\rangle\right)=B_{i,j}-\frac{1}{n-2}\left(\langle R_i, B\rangle+\langle R_j, B\rangle\right).
\end{eqnarray}
With Equation \eqref{thm1_2}, the following equation holds:
\begin{eqnarray}
&&\sum_{\{i,j\}\in \mathbb{P}(n)}
\left\{
A_{i,j}-\frac{1}{n-2}\left(\langle R_i, A\rangle+\langle R_j, A\rangle\right)
\right\}\nonumber
\\
&&=
\sum_{\{i,j\}\in \mathbb{P}(n)}
\left\{
B_{i,j}-\frac{1}{n-2}\left(\langle R_i, B\rangle+\langle R_j, B\rangle\right)
\right\}. \label{thm1_-1}
\end{eqnarray}
Here, the left side can be calculated as follows because the number of pairs including element $k$ in $\mathbb{P}(n)$ is $n-1$:
\begin{eqnarray}
    &&\sum_{\{i,j\}\in \mathbb{P}(n)}\left\{A_{i,j}-\frac{1}{n-2}\left(\langle R_i, A\rangle+\langle R_j, A\rangle\right)\right\}\nonumber\\
    &=&\sum_{\{i,j\}\in \mathbb{P}(n)}A_{i,j}-\frac{n-1}{n-2}\sum_{k=1}^n\langle R_k, A\rangle\nonumber\\
    &=&\sum_{\{i,j\}\in \mathbb{P}(n)}A_{i,j}-\frac{n-1}{n-2}\sum_{k=1}^n\sum_{l\neq k}A_{k,l}\nonumber\\
    &=&\sum_{\{i,j\}\in \mathbb{P}(n)}A_{i,j}-\frac{2(n-1)}{n-2}\sum_{\{k,l\}\in \mathbb{P}(n)}A_{k,l}\nonumber\\
    &=&-\frac{n}{n-2}\sum_{\{i,j\}\in \mathbb{P}(n)}A_{i,j}.\label{thm1_0}
\end{eqnarray}
Using Equation \eqref{thm1_0}, Equation \eqref{thm1_-1} is transformed into the following:
\begin{eqnarray}
    -\frac{n}{n-2}\sum_{\{i,j\}\in\mathbb{P}(n)}A_{i,j}=-\frac{n}{n-2}\sum_{\{i,j\}\in\mathbb{P}(n)}B_{i,j}.
\end{eqnarray}
Therefore,
\begin{eqnarray}
\sum_{\{i,j\}\in\mathbb{P}(n)}A_{i,j}=\sum_{\{i,j\}\in\mathbb{P}(n)}B_{i,j}.\label{thm1_1}
\end{eqnarray}
The following equation holds for any pairing $S$, by Equation \eqref{thm1_2}:
\begin{eqnarray}
&&\sum_{\{i,j\}\in f_{\textrm{set}}(S)}\left\{A_{i,j}-\frac{1}{n-2}\left(\langle R_i, A\rangle+\langle R_j, A\rangle\right)\right\}\nonumber\\
&&=\sum_{\{i,j\}\in f_{\textrm{set}}(S)}\left\{B_{i,j}-\frac{1}{n-2}\left(\langle R_i, B\rangle+\langle R_j, B\rangle\right)\right\}.\label{thm1_2_2}
\end{eqnarray}
Here, the following equation holds. Note that $\{i,j\}$ belongs to $f_{\textrm{set}}(S)$; hence,  $\langle R_k,A \rangle$  appears only once and all index $k$ ranging from $1$ to $n$ appear over the summation:
\begin{eqnarray}
    \sum_{\{i,j\}\in f_{\textrm{set}}(S)}\left(\langle R_i, A\rangle+\langle R_j, A\rangle\right)&=&\sum_{k=1}^n\langle R_k, A\rangle\label{thm1_2_3_pre1}\\
    &=& \sum_{k=1}^n \sum_{l, l \neq k}A_{k,l}\nonumber\\
    &=& 2\sum_{\{k,l\}\in\mathbb{P}(n)} A_{k,l}.\label{thm1_2_3}
\end{eqnarray}
For $B$, the following equation also holds:
\begin{eqnarray}
    \sum_{\{i,j\}\in f_{\textrm{set}}(S)}\left(\langle R_i, B\rangle+\langle R_j, B\rangle\right)&=&\sum_{k=1}^n\langle R_k, B\rangle\label{thm1_2_3_pre2}\\
    &=&2\sum_{\{k,l\}\in\mathbb{P}(n)} B_{k,l}. \label{thm1_2_3_2}
\end{eqnarray}
Using these transformations, Equation \eqref{thm1_2_2} is transformed as follows:
\begin{eqnarray}
\langle S,A\rangle-\frac{2}{n-2}\sum_{\{k,l\}\in\mathbb{P}(n)}A_{k,l}=\langle S,B\rangle-\frac{2}{n-2}\sum_{\{k,l\}\in\mathbb{P}(n)}B_{k,l}.
\end{eqnarray}
With Equation \eqref{thm1_1},
\begin{eqnarray}
&&\langle S,A\rangle=\langle S,B\rangle.
\end{eqnarray}
Then, $ A\sim B$ holds.

Second, we prove the necessity. 
We assume that $ A\sim B$ holds. 
We define $ A^*\in\Omega_n$ as follows:
\begin{eqnarray}
    A^*_{i,j}\equiv\frac{1}{n-2}(\langle R_i, A\rangle+\langle R_j, A\rangle)+B_{i,j}-\frac{1}{n-2}(\langle R_i, B\rangle+\langle R_j, B\rangle)\label{thm2_asta}.
\end{eqnarray}
By Equations \eqref{thm1_2_3_pre1}, \eqref{thm1_2_3_pre2} and \eqref{thm2_asta},
\begin{eqnarray}
\forall S\in\mathbb{S}(n), \langle S,A^*\rangle&=&\sum_{\{i,j\}\in f_{\textrm{set}}(S)}A^*_{i,j}\nonumber\\
&=&\langle S,B\rangle+\frac{1}{n-2}\sum_{i=1}^n\langle R_i, A\rangle-\frac{1}{n-2}\sum_{i=1}^n\langle R_i, B\rangle\label{thm2_asta2}.
\end{eqnarray}
We derive the relationship between $\sum_{i=1}^n\langle R_i, A\rangle$ and $\sum_{S\in\mathbb{S}(n)}\langle S,A\rangle$ here in order to transform Equation \eqref{thm2_asta2}. 
By Equation \eqref{thm1_2_3},
\begin{eqnarray}
    \sum_{i=1}^n\langle R_i, A\rangle=2\sum_{\{i,j\}\in\mathbb{P}(n)}A_{i,j}\label{thm1_2_4}. \label{thm1_2_5}
\end{eqnarray}
For $\sum_{S\in\mathbb{S}(n)}\langle S,A\rangle$, we focus on the fact that the number of appearances of $A_{i,j}$ is $(n-3)!!$,
\begin{eqnarray}
    \sum_{S\in\mathbb{S}(n)}\langle S,A\rangle=(n-3)!!\sum_{\{i,j\}\in\mathbb{P}(n)}A_{i,j}. \label{thm1_2_6}
\end{eqnarray}
With Equations \eqref{thm1_2_5} and \eqref{thm1_2_6}, the following relationship holds:
\begin{eqnarray}
    \sum_{i=1}^n\langle R_i, A\rangle=\frac{2}{(n-3)!!}\sum_{S\in\mathbb{S}(n)}\langle S,A\rangle \label{thm1_2_7}
\end{eqnarray}
Therefore, the following holds by $ A\sim B$ and Equation \eqref{thm1_2_7}:
\begin{eqnarray}
\sum_{i=1}^n\langle R_i, A\rangle&=&\frac{2}{(n-3)!!}\sum_{S\in\mathbb{S}(n)}\langle S,A\rangle\nonumber\\
&=&\frac{2}{(n-3)!!}\sum_{S\in\mathbb{S}(n)}\langle S,B\rangle\nonumber\\
&=&\sum_{i=1}^n\langle R_i, B\rangle\label{thm2_R_sum}.
\end{eqnarray}
By Equation \eqref{thm2_R_sum}, we can cancel the 2nd and 3rd terms of \eqref{thm2_asta2}, 
\begin{eqnarray}
    \langle S,A^*\rangle=\langle S,B\rangle.
\end{eqnarray}
In addition, $ A\sim B$ holds. Therefore, 
\begin{eqnarray}
    \forall S\in\mathbb{S}(n), \langle S,A^*\rangle=\langle S,B\rangle=\langle S,A\rangle\label{thm2_a}.
\end{eqnarray}
Additionally, the following also holds by $ A\sim B$ and Equation \eqref{thm2_R_sum}:
\begin{eqnarray}
    \sum_{j, j\neq i}A^*_{i,j}&=&\frac{n-1}{n-2}\langle R_i, A\rangle+\frac{1}{n-2}\sum_{j, j\neq i}\langle R_j, A\rangle+\sum_{j, j\neq i}B_{i,j}\nonumber\\
    &&-\frac{n-1}{n-2}\langle R_i, B\rangle-\frac{1}{n-2}\sum_{j, j\neq i}\langle R_j, B\rangle\nonumber\\
    &=&\frac{1}{n-2}\left(\sum_{j=1}^n\langle R_j, A\rangle-\sum_{j=1}^n\langle R_j, B\rangle\right)+\langle R_i, A\rangle\nonumber\\
    &=&\langle R_i, A\rangle\label{thm2_R_sum_eq}.
\end{eqnarray}
By Equation \eqref{thm2_R_sum_eq}, 
\begin{eqnarray}
    1\leq i \leq n, \langle R_i,A^*\rangle=\langle R_i, A\rangle\label{thm2_b}.
\end{eqnarray}
Therefore, by Equations \eqref{thm2_a} and \eqref{thm2_b} and Corollary 1,
\begin{eqnarray}
     A= A^* 
\end{eqnarray}
is valid. 
That is to say, the following equation holds:
\begin{eqnarray}
    \{i,j\}\in\mathbb{P}(n), A_{i,j}-\frac{1}{n-2}\left(\langle R_i, A\rangle+\langle R_j, A\rangle\right)=B_{i,j}-\frac{1}{n-2}\left(\langle R_i, B\rangle+\langle R_j, B\rangle\right).
\end{eqnarray}
\qed
\end{pro}

\subsection{Mean and Covariance}
Here we analyze statistical properties associated with the compatibility matrix and the total compatibility.

We define the mean values of compatibilities and total compatibilities as
\begin{eqnarray}
    C\in\Omega_n,\nonumber\\
    \mu_{\textrm{element}}(C)&\equiv& \frac{2}{n(n-1)}\sum_{1\leq i<j\leq n}C_{i,j},\nonumber\\
    \mu_{\textrm{sum}}(C)&\equiv& \frac{1}{(n-1)!!}\sum_{S\in\mathbb{S}(n)}\langle S,C\rangle.\nonumber
\end{eqnarray}
By Equation \eqref{thm1_2_6}, $\mu_{\textrm{sum}}(C)$ is transformed into 
\begin{eqnarray}
    \mu_{\textrm{sum}}(C)&\equiv& \frac{1}{(n-1)!!}\sum_{S\in\mathbb{S}(n)}\langle S,C\rangle\nonumber\\
    &=&\frac{1}{n-1}\sum_{1\leq i<j\leq n}C_{i,j}\nonumber\\
    &=&\frac{n}{2}\mu_{\textrm{element}}(C)\label{sum_def}
\end{eqnarray}
where $\mu_{\textrm{element}}(C)$ indicates the mean value of the elements of the compatibility matrix $C$ and $\mu_{\textrm{sum}}(C)$ is the mean of the total compatibility across all possible pairing with respect to the compatibility matrix $C$.

We define the square root of the covariance values for compatibilities and total compatibilities as follows:
\begin{eqnarray}
    &&\sigma_{\textrm{element}}(A, B)\equiv\sqrt{\sum_{1\leq i<j\leq n}\frac{2}{n(n-1)}\left(A_{i,j}-\mu_{\textrm{element}}( A)\right)\left(B_{i,j}-\mu_{\textrm{element}}( B)\right)},\nonumber\\
    &&\sigma_{\textrm{sum}}(A, B)\equiv\sqrt{\frac{1}{(n-1)!!}\sum_{S\in\mathbb{S}(n)}\left(\langle S,A\rangle-\mu_{\textrm{sum}}( A)\right)\left(\langle S,B\rangle-\mu_{\textrm{sum}}( B)\right)}.
\end{eqnarray}

Clearly, $\sigma_{\textrm{element}}^2(C,C)$ and $\sigma_{\textrm{sum}}^2(C,C)$ are variance values for compatibilities and total compatibilities when the compatibility matrix is $C$. 

Regarding $\sigma_{\textrm{sum}}(C,C)$, the following theorem holds.
\begin{thm}
\label{thm_vv}
    Let $I_n$ be the $n\times n$ identity matrix, $J_n$ the $n\times n$ matrix where all elements are $1$, and $ C\in\Omega_n, \hat{C}\equiv C-\mu_{\textrm{element}}(C)(J_n-I_n)$. Then, the following equation holds:
    \begin{eqnarray}
        \sigma_{\textrm{sum}}^{2}(C,C)
        =
        \frac{n(n-2)}{2(n-3)}\sigma_{\textrm{element}}^{2}(C,C)-\frac{1}{(n-1)(n-3)}\sum_{k=1}^n\langle R_k,\hat{C}\rangle^2\label{thmm3}.
    \end{eqnarray}
\end{thm}
\begin{pro}
By definition,
\begin{eqnarray}
    \sigma_{\textrm{sum}}^{2}(C,C)&=&\frac{1}{(n-1)!!}\sum_{S\in\mathbb{S}(n)}\left\{\langle S,C\rangle-\mu_{\textrm{sum}}(C)\right\}^2\nonumber
\end{eqnarray}
Using Equation \eqref{sum_def}, 
\begin{eqnarray}
    \sigma_{\textrm{sum}}^{2}(C,C)&=&\frac{1}{(n-1)!!}\sum_{S\in\mathbb{S}(n)}\left\{\langle S,C\rangle-\frac{n}{2}\mu_{\textrm{element}}(C)\right\}^2\label{thm3_1_1}
\end{eqnarray}
Here, the following equation holds:
\begin{eqnarray}
    \langle S,\hat{C} \rangle&=&\frac{1}{2}\langle S,\hat{C} \rangle_F\nonumber\\
    &=&\frac{1}{2}\langle S,C \rangle_F-\frac{1}{2}\mu_{\textrm{element}}(C)\langle S, J_n-I_n\rangle\nonumber\\
    &=&\frac{1}{2}\langle S,C \rangle_F-\frac{n}{2}\mu_{\textrm{element}}(C)\nonumber\\
    &=&\langle S,C \rangle-\frac{n}{2}\mu_{\textrm{element}}(C)\label{thm3_1_2}
\end{eqnarray}
Therefore, by Equations \eqref{thm3_1_1} and \eqref{thm3_1_2},
\begin{eqnarray}
    \sigma_{\textrm{sum}}^{2}(C,C)&=&\frac{1}{(n-1)!!}\sum_{S\in\mathbb{S}(n)}\left\{\langle S,C\rangle-\frac{n}{2}\mu_{\textrm{element}}(C)\right\}^2\nonumber\\
    &=&\frac{1}{(n-1)!!}\sum_{S\in\mathbb{S}(n)}\langle S,\hat{C}\rangle^2\nonumber\\
    &=&\frac{1}{(n-1)!!}\cdot(n-3)!!\sum_{\{i,j\}\in\mathbb{P}(n)}\hat{C}_{i,j}^2\nonumber\\
    &&+\frac{1}{(n-1)!!}\cdot(n-5)!!\sum_{\{i,j\}\in\mathbb{P}(n)}\sum_{\substack{\{k,l\}\in\mathbb{P}(n)\\\{k,l\}\cap\{i,j\}=\emptyset}}\hat{C}_{i,j}\hat{C}_{k,l}\nonumber\\
    &=&\frac{1}{n-1}\sum_{\{i,j\}\in\mathbb{P}(n)}\hat{C}_{i,j}^2+\frac{1}{(n-1)(n-3)}\sum_{\{i,j\}\in\mathbb{P}(n)}\sum_{\substack{\{k,l\}\in\mathbb{P}(n)\\\{k,l\}\cap\{i,j\}=\emptyset}}\hat{C}_{i,j}\hat{C}_{k,l}\nonumber\\
    &=&\frac{1}{n-1}\sum_{\{i,j\}\in\mathbb{P}(n)}\hat{C}_{i,j}^2+\frac{1}{(n-1)(n-3)}\sum_{\{i,j\}\in\mathbb{P}(n)}\hat{C}_{i,j}\sum_{\substack{\{k,l\}\in\mathbb{P}(n)\\\{k,l\}\cap\{i,j\}=\emptyset}}\hat{C}_{k,l}\label{thm3_1}.
\end{eqnarray}
Here, we focus on $\sum_{\substack{\{k,l\}\in\mathbb{P}(n)\\\{k,l\}\cap\{i,j\}=\emptyset}}\hat{C}_{k,l}$. This term is transformed as follows:
\begin{eqnarray}
    \sum_{\substack{\{k,l\}\in\mathbb{P}(n)\\\{k,l\}\cap\{i,j\}=\emptyset}}\hat{C}_{k,l}&=&\hat{C}_{i,j}+\sum_{\{k,l\}\in\mathbb{P}(n)}\hat{C}_{k,l}-\sum_{k,k\neq i}\hat{C}_{i,k}-\sum_{k,k\neq j}\hat{C}_{j,k}\nonumber\\
    &=&\hat{C}_{i,j}-\langle R_i,\hat{C}\rangle-\langle R_j, \hat{C}\rangle+\sum_{\{k,l\}\in\mathbb{P}(n)}\hat{C}_{k,l}\nonumber\\
    &=&\hat{C}_{i,j}-\langle R_i,\hat{C}\rangle-\langle R_j, \hat{C}\rangle+\sum_{\{k,l\}\in\mathbb{P}(n)}\left(C_{k,l}-\mu_{\textrm{element}}(C)\right)\nonumber\\
    &=&\hat{C}_{i,j}-\langle R_i,\hat{C}\rangle-\langle R_j, \hat{C}\rangle+\left(\sum_{\{k,l\}\in\mathbb{P}(n)}C_{k,l}\right)-\frac{n(n-1)}{2}\mu_{\textrm{element}}(C)\nonumber\\
    &=&\hat{C}_{i,j}-\langle R_i,\hat{C}\rangle-\langle R_j, \hat{C}\rangle.
\end{eqnarray}
Then, using this formula,
\begin{eqnarray}
    &&\sum_{\{i,j\}\in\mathbb{P}(n)}\hat{C}_{i,j}\sum_{\substack{\{k,l\}\in\mathbb{P}(n)\\\{k,l\}\neq\{i,j\}}}\hat{C}_{k,l}\nonumber\\
    &&=\sum_{\{i,j\}\in\mathbb{P}(n)}\hat{C}_{i,j}\left(\hat{C}_{i,j}-\langle R_i,\hat{C}\rangle-\langle R_j, \hat{C}\rangle\right)\nonumber\\
    &&=\sum_{\{i,j\}\in\mathbb{P}(n)}\hat{C}_{i,j}^2-\sum_{\{i,j\}\in\mathbb{P}(n)}\hat{C}_{i,j}\left(\langle R_i,\hat{C}\rangle+\langle R_j, \hat{C}\rangle\right)\nonumber\\
    &&=\sum_{\{i,j\}\in\mathbb{P}(n)}\hat{C}_{i,j}^2-\sum_{i=1}^n\sum_{j\neq i}\hat{C}_{i,j}\langle R_i,\hat{C}\rangle\nonumber\\
    &&=\sum_{\{i,j\}\in\mathbb{P}(n)}\hat{C}_{i,j}^2-\sum_{i=1}^n\langle R_i,\hat{C}\rangle^2\label{thm3_2}.
\end{eqnarray}
By Equations \eqref{thm3_1} and \eqref{thm3_2}, the following equation holds:
\begin{eqnarray}
    \sigma^2_{\textrm{sum}}(C,C)&=&\frac{n-2}{(n-1)(n-3)}\sum_{\{i,j\}\in\mathbb{P}(n)}\hat{C}_{i,j}^2-\frac{1}{(n-1)(n-3)}\sum_{k=1}^n\langle R_k,\hat{C}\rangle^2\nonumber\\
    &=&\frac{n(n-2)}{2(n-3)}\sigma^2_{\textrm{element}}(C,C)-\frac{1}{(n-1)(n-3)}\sum_{k=1}^n\langle R_k,\hat{C}\rangle^2.
\end{eqnarray}
Therefore, the theorem holds. \qed
\end{pro}

\section{Variance Optimization}
\label{Variance Optimization}
This section examines the performance enhancement from deriving a pairing that yields higher total compatibility by exploiting the algebraic structures identified in the previous section. 
We first show that the variance of the elements in a compatibility matrix affects the performance of the heuristic algorithm proposed in our previous study. Then we propose the transformation of a compatibility matrix to another one that minimizes the variance while ensuring that the total compatibility is maintained.

\subsection{Performance Degradation through the Observation Phase}

In our previous study \cite{fujita2022efficient}, we proposed an algorithm for recognizing the compatibilities among elements through multiple measurements of total compatibility.  
To summarize, we estimate the compatibility matrix denoted by $\tilde{C}\in\Omega_n$, which is given by 
\begin{eqnarray}
    &&C\in\Omega_n,\nonumber\\
    &&\tilde{C}_{i,j}=
    \begin{cases}
    0\textrm{\quad if\quad} 1\in\{i,j\}\\
    C_{i,j}-C_{1,i}-C_{1,j}+\frac{2}{n-2}\sum_{k=2}^nC_{1,k} \textrm{\quad otherwise}.
    \end{cases}\nonumber\\
    \label{ob}
\end{eqnarray}
This $\tilde{C}\in\Omega_n$ is one of the elements in the equivalence class. That is, $C \sim \hat{C}$ holds. 
By this property and Equation \eqref{ob}, the dimension of $\{S \}_{S\in\mathbb{S}(n)}$ is given by  
$(n-1)(n-2)/2$, which we refer to as $L_{\textrm{min}}(n)$.
This means that the number of observations required to grasp the compatibilities through an observation phase is $L_{\textrm{min}}(n)$. 
 
Indeed, our previous study proposed an observation algorithm which needs $\mathcal{O}(n^2)$ measurements. 
We have also confirmed numerically that the observation strategy provides a compatibility matrix, which is in the equivalence class of the ground-truth compatibility matrix $C^{g}$.
In the numerical studies, the elements of the ground-truth compatibility matrix, $C^{g}_{i,j}$, were specified by uniformly distributed random numbers in the range of $[0,1]$.

However, finding a pairing yielding a greater total compatibility becomes difficult based on $C^e$, including the above-mentioned $\tilde{C}$, even though $C^e$ is in the equivalence class where the ground-truth compatibility $C^g$ is included. 
In searching for a better pairing, we use a heuristic algorithm, which is named Pairing-2-opt \cite{fujita2022efficient}.  
We consider the difficulty comes from the fact that the variance of the elements of the compatibility matrix $\sigma_{\textrm{element}}^{2}(C^e,C^e)$ would be larger than those of $\sigma_{\textrm{element}}^{2}(C^g,C^g)$, which is highly likely to cause the combining algorithm to become stuck in a local minimum. 

Hence, our idea is to find a compatibility matrix $X$ which is in the same equivalence class of matrix $C$ 
\begin{eqnarray}
    \forall S\in\mathbb{S}(n), \langle S,X\rangle=\langle S,C\rangle
\end{eqnarray}
while simultaneously minimizing the variance of the elements of $\sigma_{\textrm{element}}^{2}(X,X)$.

\subsection{Transforming the Compatibility Matrix with Minimized Variance}
\label{Variance Optimization Algorithm}

We solve the following optimization problem:
\begin{eqnarray}
    \textrm{min}:&& \sigma^2_{\textrm{element}}(X,X),\nonumber\\
    \textrm{subject\,to}:&& X,C\in\Omega_n, C\textrm{\,is\,fixed},\nonumber\\
    &&X\sim C.
\end{eqnarray}
By Theorem \ref{thm_vv} and $\sigma^2_{\textrm{sum}}(X,X)=\sigma^2_{\textrm{sum}}(C,C)$, we transform this problem into the following form:
\begin{eqnarray}
    \textrm{min}:&& \sum_{k=1}^n\langle R_k,\hat{X}\rangle^2,\nonumber\\
    \textrm{subject\,to}:&& X,C\in\Omega_n, C\textrm{\,is\,fixed},\nonumber\\
    &&X\sim C,\nonumber\\
    &&\hat{X}\equiv X-\mu_{\textrm{element}}(X)(J_n-I_n).
\end{eqnarray}
The optimal solution for this problem holds because the sum of squares is minimized when all values are 0:
\begin{eqnarray}
    &&1\leq k \leq n, \langle R_k,\hat{X}\rangle=0. 
\end{eqnarray}
Hence, the following equation is derived: 
\begin{eqnarray}
    1\leq k \leq n, \langle R_k,X\rangle=(n-1)\mu_{\textrm{element}}(C).\label{VO_solution}
\end{eqnarray}
By Equation \eqref{VO_solution} and Theorem \ref{thm_equivalence}, the optimal solution is represented as follows:
\begin{eqnarray}
    &&X_{i,j}=\frac{2(n-1)}{n-2}\mu_{\textrm{element}}(C)+C_{i,j}-\frac{1}{n-2}\left(\langle R_i, C\rangle+\langle R_j, C\rangle\right).
\end{eqnarray}
Thus, the compatibility matrix with minimal variance is derived. In addition, this discussion and solution mean that the optimal-variance solution is unique with respect to the equivalence class.

\section{Simulation}
\label{simulation}

In this section, we evaluate the performance of the proposed method on the pairing optimization problem. 
There are two important points that should be clarified through the simulations. 
One is to quantitatively evaluate the performance reduction of the combining algorithm proposed in the previous study, based on the observation phase. 
The other is to demonstrate the performance enhancement due to the variance optimization discussed in Section~\ref{Variance Optimization}.

\subsection{Setting}
We configure the ground-truth compatibility matrix $C^g\in\Omega_n$ with two different distributions. The first is the uniform distribution:
\begin{eqnarray}
    \forall \{i,j\}\in\mathbb{P}, C^g_{i,j}\sim U(0,1).
    \label{setting}
\end{eqnarray}
Here, we denote the uniform distribution between 0 and 1 as $U(0,1)$. 
The second distribution is the Poisson distribution:
\begin{eqnarray}
    \forall \{i,j\}\in\mathbb{P}, C^g_{i,j}\sim Poisson(1).
    \label{setting2}
\end{eqnarray}
Here, we denote the Poisson distribution whose mean is $\lambda$ as $Poisson(\lambda)$. 
In the numerical simulation, the number of elements in the system $n$ varied from $100$ to $1000$ in intervals of $100$.
For each $n$, we conducted 100 trials with different randomly generated ground-truth compatibility matrices $C^g$ based on Equations \eqref{setting} or \eqref{setting2}. 
We quantified the performance for each derived pairing $S\in\mathbb{S}$ by $2\langle S,C^g\rangle/n$
and evaluated its average over 100 trials for each value of $n$.

\subsection{Simulation Flow}
The ground-truth compatibility matrix $C^g$ is transformed into $C^{e_1}$ by the observation algorithm based on Equation \eqref{ob}. 
The variance optimization transforms $C^{e_1}$ into $C^{e_2}$. 
The combining algorithm, which is called PNN+P2-opt \cite{fujita2022efficient}, yields a pairing with the intention of achieving higher total compatibility. 
The exchange limit $l$ is an internal parameter in PNN+P2-opt. This determines the number of maximum trials, and is set to $600$ in the present study. 

\begin{figure}[h]
 \centering
 \includegraphics[width=9cm]{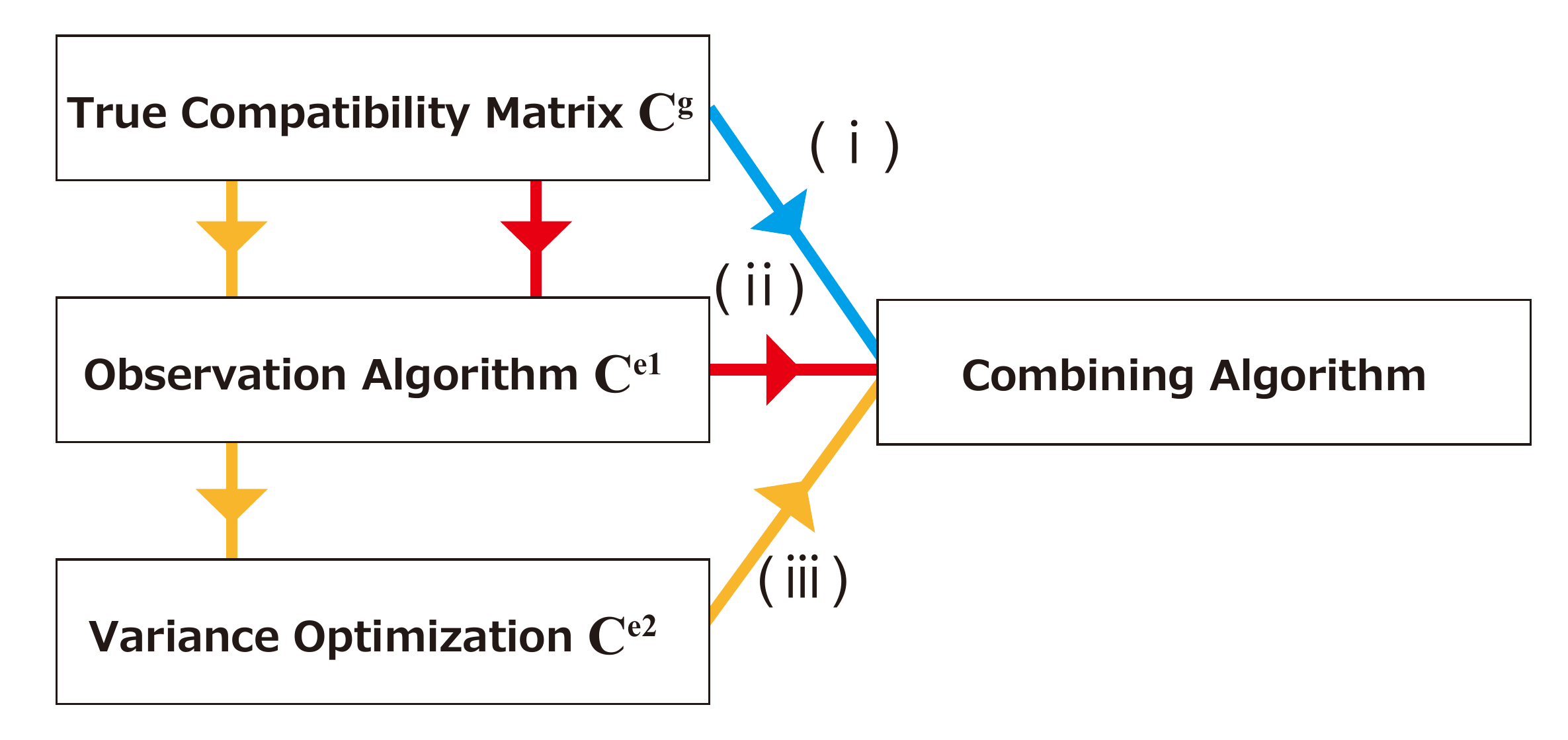}
 \caption{Schematic illustration of the three heuristic pairing optimization algorithms tested in the simulation. 
 Case (i) (blue) applies the combining algorithm directly to the ground-truth compatibility matrix $C^g$. 
 Case (ii) (red) first applies the observation algorithm to obtain an estimated compatibility matrix $C^{e_1}$, followed by the combining algorithm. 
 Case (iii) (yellow) first estimates the compatibility from observation ($C^{e_1}$), followed by the variance optimization ($C^{e_2}$), and then executes the combining algorithm.}
 \label{Flow}
\end{figure}

We evaluated the performance on the basis of $C^g$, $C^{e_1}$, and $C^{e_2}$, as shown in flows (i), (ii), and (iii), respectively, in Figure~ \ref{Flow}.

\subsection{Performance}

\begin{figure}[h]
  \begin{minipage}[b]{0.5\linewidth}
    \centering
    \includegraphics[width=7cm]{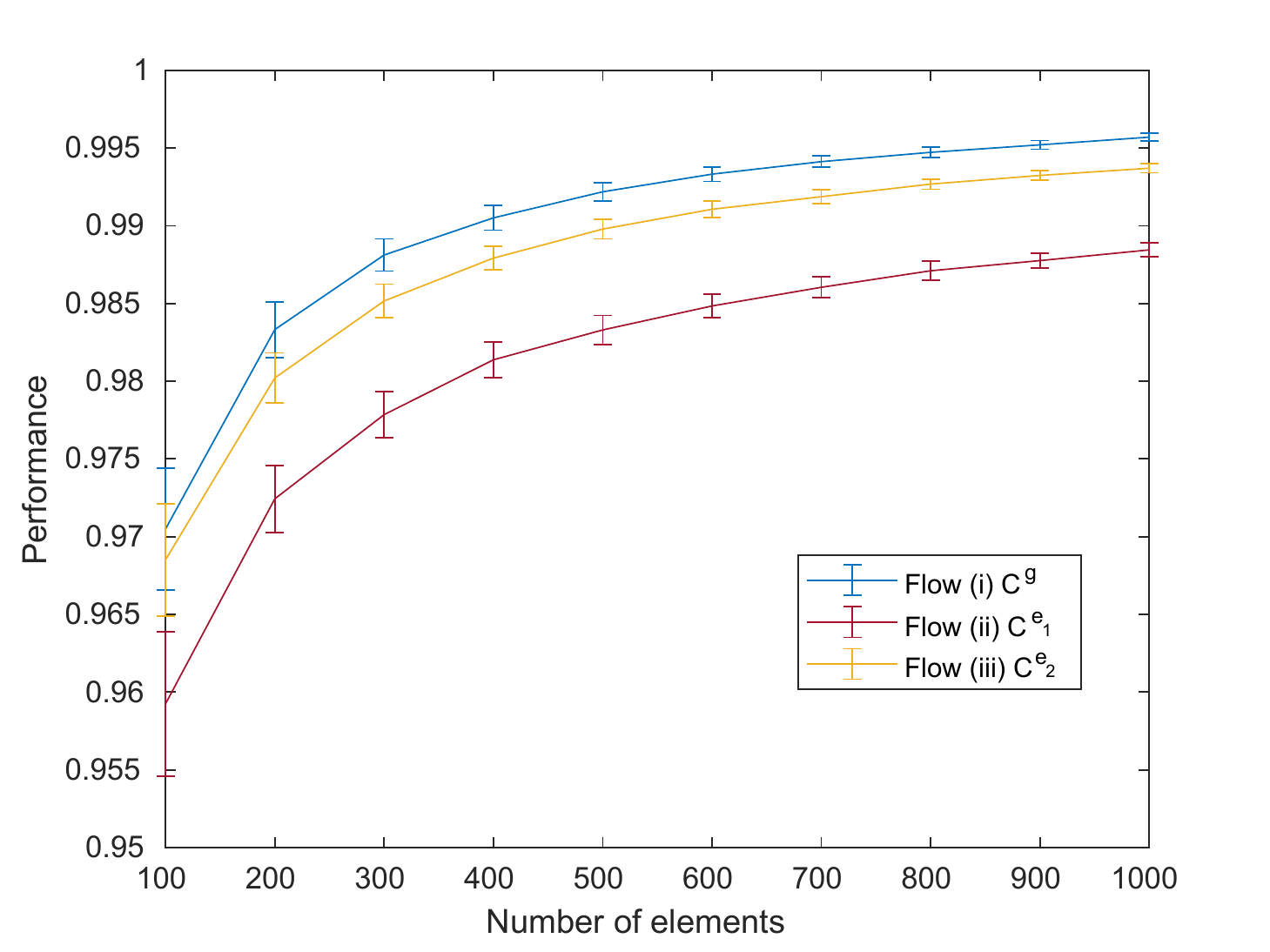}
    \subcaption{Uniform distribution}\label{Performance_1}
  \end{minipage}
  \begin{minipage}[b]{0.5\linewidth}
    \centering
    \includegraphics[width=7cm]{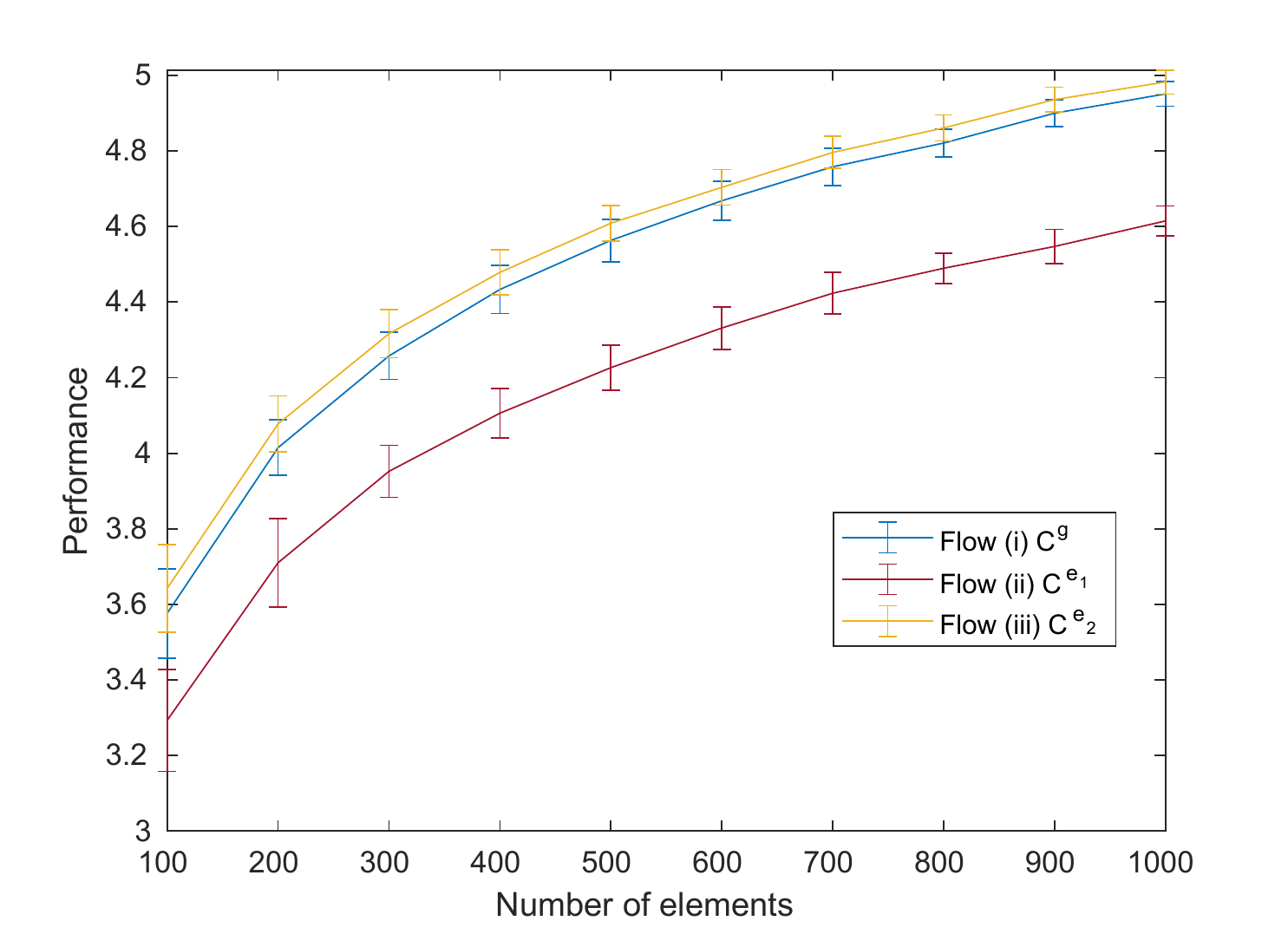}
    \subcaption{Poisson distribution}\label{Performance_2}
  \end{minipage}
  \caption{Comparison of the achieved total compatibility for Flows (i), (ii), and (iii), as described in the caption for Figure~\ref{Flow}. 
 Each graph shows the mean and standard deviation of the performance of 100 different compatibility matrices with each given number of elements, simulated under (a) uniform and (b) Poisson distributions.
 }\label{Performance}
\end{figure}

The blue, red, and yellow curves in Figure \ref{Performance} demonstrate the performance of cases (i), (ii), and (iii), respectively, as a function of the number of elements for the uniform distribution (Figure~\ref{Performance_1}) and the Poisson distribution (Figure~\ref{Performance_2}).
For the uniform distributed ground-truth we observe that the performance of case (ii) is inferior to that of case (i), demonstrating the performance degradation by the transformation from $C^g$ to $C^{e_1}$ through observation. 
Furthermore, the performance of case (iii) is enhanced compared with that of case (ii), which confirms the performance gain from variance optimization. 
The results differ for the Poisson distribution. Here, the performance of case (iii) is higher than case (i).
That is, for the Poisson case the variance optimization (Flow (iii)) not only counter-acted the performance loss of the observation algorithm (Flow (ii)), but actually enhanced the performance compared to the ground truth matrix $C^g$ (Flow (i)). 
Further numerical tests revealed that the relationship of performances for a Gaussian distribution are similar to those for the uniform distribution. 
Conversely, the performance for a binary distribution hardly differed between any of the algorithms.

\begin{figure}[h]
  \begin{minipage}[b]{0.5\linewidth}
    \centering
    \includegraphics[width=7cm]{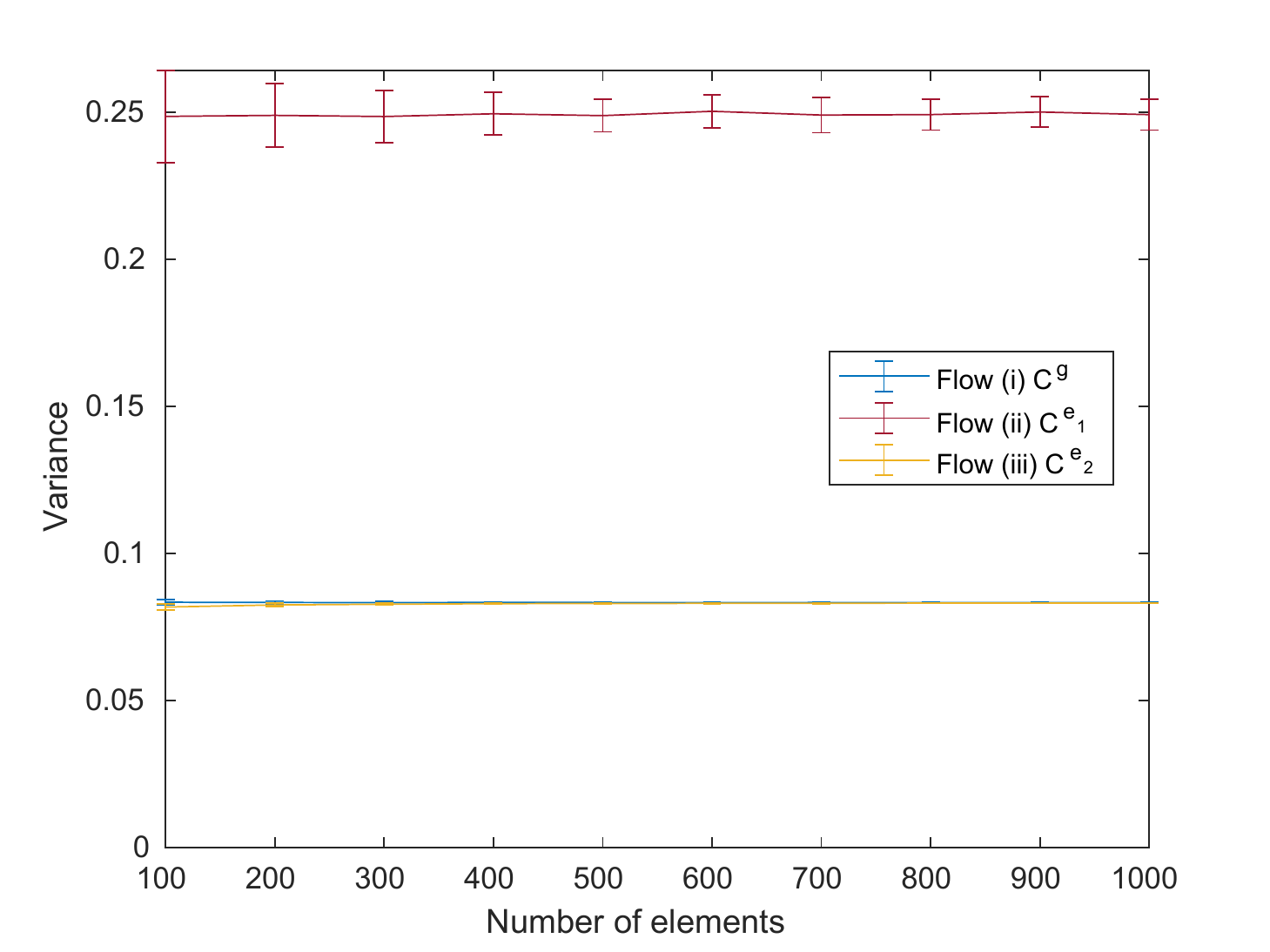}
    \subcaption{Uniform distribution}\label{Variance_1}
  \end{minipage}
  \begin{minipage}[b]{0.5\linewidth}
    \centering
    \includegraphics[width=7cm]{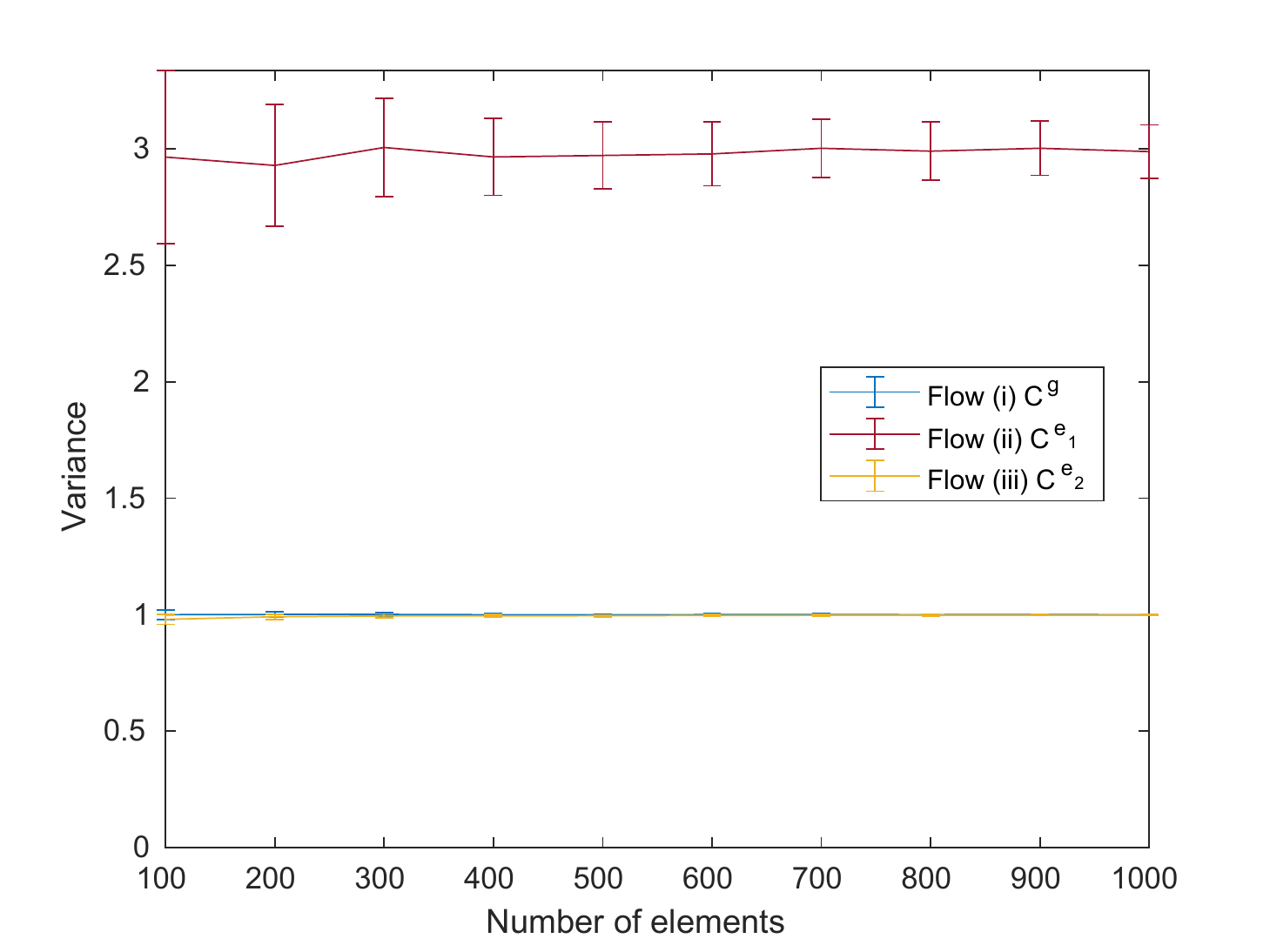}
    \subcaption{Poisson distribution}\label{Variance_2}
  \end{minipage}
  \caption{Comparison of the variance of the compatibility matrices of $C^g$, $C^{e_1}$, $C^{e_2}$ as a function of the number of elements in the system under (a) uniform and (b) Poisson distributions. 
}\label{Variance}
\end{figure}

The variance of $C^g$, $C^{e_1}$, and $C^{e_2}$ are evaluated as shown in Figure \ref{Variance} as a function of the number of elements. 
We clearly observe that the variance of $C^{e_1}$ is higher than $C^g$ while the variance of $C^{e_2}$ becomes comparable to the ground-truth case $C^g$ for both the uniform and Poisson distributions. 

From these numerical results, we can conclude that the variance optimization minimizes the variance and enhances the performance of the achieved total compatibility. 
It is worth noting that the performance with the uniform distribution after variance optimization is still lower than the case based on the ground-truth matrix $C^g$, as observed in Figure \ref{Performance_1}.  
This occurs because the variance optimization algorithm does not transform $C^{e_1}$ to the original compatibility matrix $C^g$. 
In other words, there exist additional factors that influence the performance of the combining algorithm that are related to the compatibility distribution. The distribution of the original compatibility $C^g$ (uniform distribution) is seemingly beneficial for the performance of the heuristic combining algorithm, even when compared to the compatibility matrix with minimum variance $C^{e_2}$.

\section{Conclusion}
\label{Conclusion}
One of the most challenging issues in the pairing problem is how to understand the underlying compatibilities among the elements under study. 
An accurate and efficient approach is essential for practical applications such as wireless communications and online social networks.
This study reveals several algebraic structures in the pairing optimization problem. 

We introduce an equivalence class in the compatibility matrices, containing matrices that yield the same total compatibility although the matrices themselves differ. 
This can also be expressed through a conserved value or invariance in the equivalence class. 
Based on such insights, we propose a transformation of the initially estimated compatibility matrix to another form that minimizes the variance of the elements. 
We demonstrate that the highest total compatibility found heuristically is improved significantly with the proposed transformation relative to the direct approach.

\appendix
\section[\appendixname~\thesection]{Matrix Form of Conserved Quantities}
\label{Matrix Form of Conserved Quantities}
In Theorem \ref{thm_equivalence}, the following values are conserved in the same equivalence class.
\begin{eqnarray}
    \forall \{i,j\}\in\mathbb{P}(n), C_{i,j}-\frac{1}{n-2}\left(\langle R_i, C\rangle+\langle R_j, C\rangle\right).\label{conserved}
\end{eqnarray} 
We can transform Equation \eqref{conserved} into the following form using the Hadamard product $\circ$.
\begin{eqnarray}
    C-\frac{1}{n-2}(J_n-I_n)\circ (J_nC+CJ_n).
\end{eqnarray}
Therefore, the following equation holds:
\begin{eqnarray}
    &&A\sim B\textrm{\quad if and only if \quad} \nonumber\\
    &&A-\frac{1}{n-2}(J_n-I_n)\circ (J_nA+AJ_n)=B-\frac{1}{n-2}(J_n-I_n)\circ (J_nB+BJ_n).
\end{eqnarray}

\bibliographystyle{plain}
\bibliography{sample.bib}

\begin{thebibliography}{10}

\bibitem{aldababsa2018tutorial}
Mahmoud Aldababsa, Mesut Toka, Selahattin G{\"o}k{\c{c}}eli,
  G{\"u}ne{\c{s}}~Karabulut Kurt, and O{\u{g}}uz Kucur.
\newblock A tutorial on nonorthogonal multiple access for {5G} and beyond.
\newblock {\em Wireless communications and mobile computing}, 2018, 2018.

\bibitem{ali2021optimizing}
Zain Ali, Wali~Ullah Khan, Asim Ihsan, Omer Waqar, Guftaar Ahmad~Sardar Sidhu,
  and Neeraj Kumar.
\newblock Optimizing resource allocation for 6g {NOMA}-enabled cooperative
  vehicular networks.
\newblock {\em IEEE Open Journal of Intelligent Transportation Systems},
  2:269--281, 2021.

\bibitem{bellur2007improved}
Umesh Bellur and Roshan Kulkarni.
\newblock Improved matchmaking algorithm for semantic web services based on
  bipartite graph matching.
\newblock In {\em IEEE international conference on web services (ICWS 2007)},
  pages 86--93. IEEE, 2007.

\bibitem{chen2019proportional}
Liang Chen, Lin Ma, and Yubin Xu.
\newblock Proportional fairness-based user pairing and power allocation
  algorithm for non-orthogonal multiple access system.
\newblock {\em IEEE Access}, 7:19602--19615, 2019.

\bibitem{cygan2015algorithmic}
Marek Cygan, Harold~N Gabow, and Piotr Sankowski.
\newblock Algorithmic applications of baur-strassen’s theorem: Shortest
  cycles, diameter, and matchings.
\newblock {\em Journal of the ACM (JACM)}, 62(4):1--30, 2015.

\bibitem{ding2015impact}
Zhiguo Ding, Pingzhi Fan, and H~Vincent Poor.
\newblock Impact of user pairing on {5G} nonorthogonal multiple-access downlink
  transmissions.
\newblock {\em IEEE Transactions on Vehicular Technology}, 65(8):6010--6023,
  2015.

\bibitem{duan2010approximating}
Ran Duan and Seth Pettie.
\newblock Approximating maximum weight matching in near-linear time.
\newblock In {\em 2010 IEEE 51st Annual Symposium on Foundations of Computer
  Science}, pages 673--682. IEEE, 2010.

\bibitem{duan2014linear}
Ran Duan and Seth Pettie.
\newblock Linear-time approximation for maximum weight matching.
\newblock {\em Journal of the ACM (JACM)}, 61(1):1--23, 2014.

\bibitem{edmonds1965paths}
Jack Edmonds.
\newblock Paths, trees, and flowers.
\newblock {\em Canadian Journal of mathematics}, 17:449--467, 1965.

\bibitem{ergin2017dual}
Haluk Ergin, Tayfun S{\"o}nmez, and M~Utku {\"U}nver.
\newblock Dual-donor organ exchange.
\newblock {\em Econometrica}, 85(5):1645--1671, 2017.

\bibitem{fujita2022efficient}
Naoki Fujita, Nicolas Chauvet, Andr{\'e} R{\"o}hm, Ryoichi Horisaki, Aohan Li,
  Mikio Hasegawa, and Makoto Naruse.
\newblock Efficient pairing in unknown environments: Minimal observations and
  tsp-based optimization.
\newblock {\em IEEE Access}, 10:57630--57640, 2022.

\bibitem{gabow1990data}
Harold~N Gabow.
\newblock Data structures for weighted matching and nearest common ancestors
  with linking.
\newblock In {\em Proceedings of the first annual ACM-SIAM symposium on
  Discrete algorithms}, pages 434--443, 1990.

\bibitem{gale1962college}
David Gale and Lloyd~S Shapley.
\newblock College admissions and the stability of marriage.
\newblock {\em The American Mathematical Monthly}, 69(1):9--15, 1962.

\bibitem{gambetta2017building}
Jay~M Gambetta, Jerry~M Chow, and Matthias Steffen.
\newblock Building logical qubits in a superconducting quantum computing
  system.
\newblock {\em npj quantum information}, 3(1):1--7, 2017.

\bibitem{gao20113d}
Yue Gao, Qionghai Dai, Meng Wang, and Naiyao Zhang.
\newblock {3D} model retrieval using weighted bipartite graph matching.
\newblock {\em Signal Processing: Image Communication}, 26(1):39--47, 2011.

\bibitem{halim2019combinatorial}
A~Hanif Halim and IJAoCMiE Ismail.
\newblock Combinatorial optimization: comparison of heuristic algorithms in
  travelling salesman problem.
\newblock {\em Archives of Computational Methods in Engineering},
  26(2):367--380, 2019.

\bibitem{hanke2010new}
Sven Hanke and Stefan Hougardy.
\newblock {\em New approximation algorithms for the weighted matching problem}.
\newblock Citeseer, 2010.

\bibitem{higuchi2015non}
Kenichi Higuchi and Anass Benjebbour.
\newblock Non-orthogonal multiple access ({NOMA}) with successive interference
  cancellation for future radio access.
\newblock {\em IEICE Transactions on Communications}, 98(3):403--414, 2015.

\bibitem{huang2012efficient}
Chien-Chung Huang and Telikepalli Kavitha.
\newblock Efficient algorithms for maximum weight matchings in general graphs
  with small edge weights.
\newblock In {\em Proceedings of the Twenty-Third Annual ACM-SIAM Symposium on
  Discrete Algorithms}, pages 1400--1412. SIAM, 2012.

\bibitem{kohl2004airline}
Niklas Kohl and Stefan~E Karisch.
\newblock Airline crew rostering: Problem types, modeling, and optimization.
\newblock {\em Annals of Operations Research}, 127(1):223--257, 2004.

\bibitem{pettie2012simple}
Seth Pettie.
\newblock A simple reduction from maximum weight matching to maximum
  cardinality matching.
\newblock {\em Information Processing Letters}, 112(23):893--898, 2012.

\bibitem{roth1982economics}
Alvin~E Roth.
\newblock The economics of matching: Stability and incentives.
\newblock {\em Mathematics of operations research}, 7(4):617--628, 1982.

\bibitem{shahab2016user}
Muhammad~Basit Shahab, Mohammad Irfan, Md~Fazlul Kader, and Soo Young~Shin.
\newblock User pairing schemes for capacity maximization in non-orthogonal
  multiple access systems.
\newblock {\em Wireless Communications and Mobile Computing},
  16(17):2884--2894, 2016.

\bibitem{zhang2020energy}
Haijun Zhang, Yanan Duan, Keping Long, and Victor~CM Leung.
\newblock Energy efficient resource allocation in terahertz downlink {NOMA}
  systems.
\newblock {\em IEEE Transactions on Communications}, 69(2):1375--1384, 2020.

\bibitem{zhu2018optimal}
Lipeng Zhu, Jun Zhang, Zhenyu Xiao, Xianbin Cao, and Dapeng~Oliver Wu.
\newblock Optimal user pairing for downlink non-orthogonal multiple access
  ({NOMA}).
\newblock {\em IEEE Wireless Communications Letters}, 8(2):328--331, 2018.

\end{thebibliography}
\end{document}